\documentclass{aa}
\usepackage{graphicx,txfonts}
\graphicspath{{./fig/}{./png/}}
\usepackage[normalem]{ulem}
\begin{document}

%
\newcommand{\EQ}{\begin{equation}}
\newcommand{\EN}{\end{equation}}
\newcommand{\EQA}{\begin{eqnarray}}
\newcommand{\ENA}{\end{eqnarray}}
\newcommand{\eq}[1]{(\ref{#1})}
\newcommand{\EEq}[1]{Equation~(\ref{#1})}
\newcommand{\Eq}[1]{Eq.~(\ref{#1})}
\newcommand{\Eqs}[2]{Eqs~(\ref{#1}) and~(\ref{#2})}
\newcommand{\EEqs}[2]{Equations~(\ref{#1}) and~(\ref{#2})}
\newcommand{\Eqss}[2]{Eqs~(\ref{#1})--(\ref{#2})}
\newcommand{\eqs}[2]{(\ref{#1}) and~(\ref{#2})}
\newcommand{\eqss}[2]{(\ref{#1})--(\ref{#2})}
\newcommand{\App}[1]{Appendix~\ref{#1}}
\newcommand{\Sec}[1]{Sect.~\ref{#1}}
\newcommand{\Secs}[2]{Sects~\ref{#1} and \ref{#2}}
\newcommand{\Secss}[2]{Sects~\ref{#1}--\ref{#2}}
\newcommand{\Fig}[1]{Fig.~\ref{#1}}
\newcommand{\FFig}[1]{Figure~\ref{#1}}
\newcommand{\Tab}[1]{Table~\ref{#1}}
\newcommand{\Figs}[2]{Figs~\ref{#1} and \ref{#2}}
\newcommand{\FFigs}[2]{Figs~\ref{#1}--\ref{#2}}
\newcommand{\Tabs}[2]{Tables~\ref{#1} and \ref{#2}}
\newcommand{\bra}[1]{\langle #1\rangle}
\newcommand{\bbra}[1]{\left\langle #1\right\rangle}
\newcommand{\ko}{k_0}
\newcommand{\mean}[1]{\overline{ #1}}
\newcommand{\meanF}{\overline{\cal F}}
\newcommand{\meanEMF}{\vec{\cal E}}
\newcommand{\meanemf}{{\cal E}}
\newcommand{\meanFF}{\overline{\mbox{\boldmath ${\cal F}$}} {}}
\newcommand{\meanA}{\overline{A}}
\newcommand{\meanB}{\overline{B}}
\newcommand{\meanC}{\overline{C}}
\newcommand{\meanJ}{\overline{J}}
\newcommand{\meanU}{\overline{U}}
\newcommand{\meanW}{\overline{W}}
\newcommand{\meanO}{\overline{\Omega}}
\newcommand{\meanc}{\overline{c}}
\newcommand{\meanu}{\overline{u}}
\newcommand{\meanT}{\overline{T}}
\newcommand{\meanAA}{\overline{\vec{A}}}
\newcommand{\meanBB}{\overline{\vec{B}}}
\newcommand{\meanJJ}{\overline{\vec{J}}}
\newcommand{\meanUU}{\overline{\vec{U}}}
\newcommand{\meanWW}{\overline{\vec{W}}}
\newcommand{\meanQQ}{\overline{\vec{Q}}}
\newcommand{\meanTT}{\overline{\vec{T}}}
\newcommand{\meanEE}{\overline{\vec{E}}}
\newcommand{\meanBBp}{\overline{\vec{B}}_{\rm p}}
\newcommand{\meanUUp}{\overline{\vec{U}}_{\rm p}}
\newcommand{\meanuu}{\overline{\mbox{\boldmath $u$}}{}}{}
\newcommand{\meanoo}{\overline{\mbox{\boldmath $\omega$}}{}}{}
\newcommand{\meanuxB}{\overline{\mbox{\boldmath $\delta u\times \delta B$}}{}}{}
\newcommand{\meanJB}{\overline{\mbox{\boldmath $J\cdot B$}}{}}{}
\newcommand{\meanAB}{\overline{\mbox{\boldmath $A\cdot B$}}{}}{}
\newcommand{\meanjb}{\overline{\mbox{\boldmath $j\cdot b$}}{}}{}
%
%
\newcommand{\teps}{\tilde{\epsilon} {}}
%
%
\newcommand{\eee}{\hat{\mbox{\boldmath $e$}} {}}
\newcommand{\nnn}{\hat{\mbox{\boldmath $n$}} {}}
\newcommand{\vvv}{\hat{\mbox{\boldmath $v$}} {}}
\newcommand{\rr}{\hat{\mbox{\boldmath $r$}} {}}
\newcommand{\xxx}{\hat{\mbox{\boldmath $x$}} {}}
\newcommand{\yyy}{\hat{\mbox{\boldmath $y$}} {}}
\newcommand{\zz}{\hat{\mbox{\boldmath $z$}} {}}
\newcommand{\pp}{\hat{\vec{\phi}}}
\newcommand{\ttt}{\hat{\mbox{\boldmath $\theta$}} {}}
\newcommand{\OOO}{\hat{\mbox{\boldmath $\Omega$}} {}}
\newcommand{\ooo}{\hat{\mbox{\boldmath $\omega$}} {}}
\newcommand{\BBBB}{\hat{\mbox{\boldmath $B$}} {}}
\newcommand{\Bhat}{\hat{B}}
%
%
\newcommand{\gggg}{\mbox{\boldmath $g$} {}}
\newcommand{\ddd}{\mbox{\boldmath $d$} {}}
\newcommand{\rrr}{\mbox{\boldmath $r$} {}}
\newcommand{\xx}{{\vec{x}}}
\newcommand{\yy}{\mbox{\boldmath $y$} {}}
\newcommand{\zzz}{\mbox{\boldmath $z$} {}}
\newcommand{\vv}{\mbox{\boldmath $v$} {}}
\newcommand{\ww}{\mbox{\boldmath $w$} {}}
\newcommand{\mm}{\mbox{\boldmath $m$} {}}
\newcommand{\PP}{\mbox{\boldmath $P$} {}}
\newcommand{\bp}{\mbox{\boldmath $p$} {}}
\newcommand{\II}{\mbox{\boldmath $I$} {}}
\newcommand{\qq}{{\vec{q}}}
\newcommand{\QQ}{{\vec{Q}}}
\newcommand{\UU}{{\vec{U}}}
\newcommand{\uu}{{\vec{u}}}
\newcommand{\BB}{{\vec{B}}}
\newcommand{\JJ}{{\vec{J}}}
\newcommand{\jj}{{\vec{j}}}
\newcommand{\AAA}{{\vec{A}}}
\newcommand{\aaaa}{{\vec{a}}}
\newcommand{\bb}{{\vec{b}}}
\newcommand{\cc}{{\vec{c}}}
\newcommand{\nn}{\mbox{\boldmath $n$} {}}
\newcommand{\ee}{\mbox{\boldmath $e$} {}}
\newcommand{\ff}{\mbox{\boldmath $f$} {}}
\newcommand{\hh}{\mbox{\boldmath $h$} {}}
\newcommand{\EE}{{\vec{E}}}
\newcommand{\FF}{{\vec{F}}}
\newcommand{\GG}{{\vec{G}}}
\newcommand{\HH}{{\vec{H}}}
\newcommand{\TT}{\mbox{\boldmath $T$} {}}
\newcommand{\CC}{\mbox{\boldmath $C$} {}}
\newcommand{\KK}{\mbox{\boldmath $K$} {}}
\newcommand{\MM}{\mbox{\boldmath $M$} {}}
\newcommand{\kk}{{\vec{k}}}
\newcommand{\SSS}{{\vec{S}}}
\newcommand{\grav}{\mbox{\boldmath $g$} {}}
\newcommand{\nab}{\vec{\nabla}}
\newcommand{\OO}{\mbox{\boldmath $\Omega$} {}}
\newcommand{\oo}{\vec{\omega}}
\newcommand{\ddelta}{\vec{\delta}}
\newcommand{\ttau}{\mbox{\boldmath $\tau$} {}}
\newcommand{\LLambda}{\mbox{\boldmath $\Lambda$} {}}
\newcommand{\llambda}{\mbox{\boldmath $\lambda$} {}}
\newcommand{\pomega}{\mbox{\boldmath $\varpi$} {}}
%
%
\newcommand{\PPPP}{\mbox{\boldmath ${\sf P}$} {}}
\newcommand{\MMMM}{\mbox{\boldmath ${\sf M}$} {}}
\newcommand{\SSSS}{\mbox{\boldmath ${\sf S}$} {}}
\newcommand{\GGGG}{\mbox{\boldmath ${\sf G}$} {}}
\newcommand{\IIII}{\mbox{\boldmath ${\sf I}$} {}}
\newcommand{\LLLL}{\mbox{\boldmath ${\sf L}$} {}}
\newcommand{\RRRR}{\mbox{\boldmath ${\sf R}$} {}}
\newcommand{\BBB}{\mbox{\boldmath ${\cal B}$} {}}
\newcommand{\emf}{\mbox{\boldmath ${\cal E}$} {}}
\newcommand{\FFF}{\mbox{\boldmath ${\cal F}$} {}}
\newcommand{\GGG}{\mbox{\boldmath ${\cal G}$} {}}
\newcommand{\HHH}{\mbox{\boldmath ${\cal H}$} {}}
\newcommand{\QQQ}{\mbox{\boldmath ${\cal Q}$} {}}
%
%
\newcommand{\e}{{\rm e}}
\newcommand{\ii}{{\rm i}}
\newcommand{\grad}{{\rm grad} \, {}}
\newcommand{\curl}{{\rm curl} \, {}}
\newcommand{\dive}{{\rm div}  \, {}}
\newcommand{\Dive}{{\rm Div}  \, {}}
\newcommand{\sgn}{{\rm sgn}  \, {}}
\newcommand{\DD}{{\rm D} {}}
\newcommand{\DDD}{{\cal D} {}}
\newcommand{\dd}{{\rm d} {}}
\newcommand{\const}{{\rm const}  {}}
\def\degr{\hbox{$^\circ$}}
\def\la{\mathrel{\mathchoice {\vcenter{\offinterlineskip\halign{\hfil
$\displaystyle##$\hfil\cr<\cr\sim\cr}}}
{\vcenter{\offinterlineskip\halign{\hfil$\textstyle##$\hfil\cr<\cr\sim\cr}}}
{\vcenter{\offinterlineskip\halign{\hfil$\scriptstyle##$\hfil\cr<\cr\sim\cr}}}
{\vcenter{\offinterlineskip\halign{\hfil$\scriptscriptstyle##$\hfil\cr<\cr\sim\cr}}}}}
\def\ga{\mathrel{\mathchoice {\vcenter{\offinterlineskip\halign{\hfil
$\displaystyle##$\hfil\cr>\cr\sim\cr}}}
{\vcenter{\offinterlineskip\halign{\hfil$\textstyle##$\hfil\cr>\cr\sim\cr}}}
{\vcenter{\offinterlineskip\halign{\hfil$\scriptstyle##$\hfil\cr>\cr\sim\cr}}}
{\vcenter{\offinterlineskip\halign{\hfil$\scriptscriptstyle##$\hfil\cr>\cr\sim\cr}}}}}
%
%
\def\Ta{\mbox{\rm Ta}}
\def\Ra{\mbox{\rm Ra}}
\def\Ma{\mbox{\rm Ma}}
\def\Roo{\mbox{\rm Ro}^{-1}}
\def\Pra{\mbox{\rm Pr}}
\def\Pran{\mbox{\rm Pr}}
\def\Pm{\mbox{\rm Pr}_\mathrm{M}}
\def\Rm{R_\mathrm{m}}
\def\Rey{\mbox{\rm Re}}
\def\Pe{\mbox{\rm Pe}}
\def\cs{c_\mathrm{s}}
\def\sigmaSB{\sigma_{\rm SB}}
\newcommand{\ea}{{\rm et al.\ }}
\newcommand{\eaa}{{\rm et al.\ }}
\def\half{{\textstyle{1\over2}}}
\def\threehalf{{\textstyle{3\over2}}}
\def\onethird{{\textstyle{1\over3}}}
\def\onesixth{{\textstyle{1\over6}}}
\def\twothird{{\textstyle{2\over3}}}
\def\fourthird{{\textstyle{4\over3}}}
\def\quarter{{\textstyle{1\over4}}}
\newcommand{\W}{\,{\rm W}}
\newcommand{\V}{\,{\rm V}}
\newcommand{\kV}{\,{\rm kV}}
\newcommand{\T}{\,{\rm T}}
\newcommand{\G}{\,{\rm G}}
\newcommand{\Gtwopers}{\,{\rm G^2\!/s}}
\newcommand{\Hz}{\,{\rm Hz}}
\newcommand{\nHz}{\,{\rm nHz}}
\newcommand{\kHz}{\,{\rm kHz}}
\newcommand{\kG}{\,{\rm kG}}
\newcommand{\K}{\,{\rm K}}
\newcommand{\C}{\,{\rm C}}
\newcommand{\g}{\,{\rm g}}
\newcommand{\s}{\,{\rm s}}
\newcommand{\mpers}{\,{\rm m/s}}
\newcommand{\ms}{\,{\rm ms}}
\newcommand{\cm}{\,{\rm cm}}
\newcommand{\m}{\,{\rm m}}
\newcommand{\km}{\,{\rm km}}
\newcommand{\kms}{\,{\rm km\s^{-1}}}
\newcommand{\kg}{\,{\rm kg}}
\newcommand{\kW}{\,{\rm kW}}
\newcommand{\MW}{\,{\rm MW}}
\newcommand{\Mm}{\,{\rm Mm}}
\newcommand{\Mx}{\,{\rm Mx}}
\newcommand{\pc}{\,{\rm pc}}
\newcommand{\kpc}{\,{\rm kpc}}
\newcommand{\Mpc}{\,{\rm Mpc}}
\newcommand{\yr}{\,{\rm yr}}
\newcommand{\Myr}{\,{\rm Myr}}
\newcommand{\Gyr}{\,{\rm Gyr}}
\newcommand{\erg}{\,{\rm erg}}
\newcommand{\mol}{\,{\rm mol}}
\newcommand{\dyn}{\,{\rm dyn}}
\newcommand{\days}{\,{\rm d}}
\newcommand{\J}{\,{\rm J}}
\newcommand{\RM}{\,{\rm RM}}
\newcommand{\EM}{\,{\rm EM}}
\newcommand{\AU}{\,{\rm AU}}
\newcommand{\A}{\,{\rm A}}
%
%
\newcommand{\yastroph}[2]{ #1, astro-ph/#2}
\newcommand{\yjas}[3]{ #1, {J. Atmosph. Sci.,} {#2}, #3}
\newcommand{\ycsf}[3]{ #1, {Chaos, Solitons \& Fractals,} {#2}, #3}
\newcommand{\yepl}[3]{ #1, {Europhys. Lett.,} {#2}, #3}
\newcommand{\yaj}[3]{ #1, {AJ,} {#2}, #3}
\newcommand{\yjgr}[3]{ #1, {JGR,} {#2}, #3}
\newcommand{\ysol}[3]{ #1, {Sol. Phys.,} {#2}, #3}
\newcommand{\ysph}[3]{ #1, {Sol. Phys.,} {#2}, #3}
\newcommand{\yapj}[3]{ #1, {ApJ,} {#2}, #3}
\newcommand{\yapjl}[3]{ #1, {ApJ,} {#2}, #3}
\newcommand{\yapjs}[3]{ #1, {ApJ Suppl.,} {#2}, #3}
\newcommand{\yan}[3]{ #1, {AN,} {#2}, #3}
\newcommand{\ymhd}[3]{ #1, {Mag\-ne\-to\-hydro\-dyn.} {#2}, #3}
\newcommand{\ypnas}[3]{ #1, {PNAS,} {#2}, #3}
\newcommand{\yana}[3]{ #1, {A\&A,} {#2}, #3}
\newcommand{\yanas}[3]{ #1, {A\&AS,} {#2}, #3}
\newcommand{\yanar}[3]{ #1, {A\&AR,} {#2}, #3}
\newcommand{\yass}[3]{ #1, {Ap\&SS,} {#2}, #3}
\newcommand{\ygafd}[3]{ #1, {GApFD,} {#2}, #3}
\newcommand{\ypasj}[3]{ #1, {Publ. Astron. Soc. Japan,} {#2}, #3}
\newcommand{\yjfm}[3]{ #1, {JFM,} {#2}, #3}
\newcommand{\ypf}[3]{ #1, {PhFl,} {#2}, #3}
\newcommand{\ypp}[3]{ #1, {PhPl,} {#2}, #3}
\newcommand{\ysov}[3]{ #1, {Sov. Astron.,} {#2}, #3}
\newcommand{\ysovl}[3]{ #1, {Sov. Astron. Lett.,} {#2}, #3}
\newcommand{\yjetp}[3]{ #1, {Sov. Phys. JETP,} {#2}, #3}
\newcommand{\yphy}[3]{ #1, {Physica,} {#2}, #3}
\newcommand{\yannr}[3]{ #1, {ARA\&A,} {#2}, #3}
\newcommand{\yprs}[3]{ #1, {Proc. Roy. Soc. Lond.,} {#2}, #3}
\newcommand{\yprl}[3]{ #1, {PRL,} {#2}, #3}
\newcommand{\ypre}[3]{ #1, {PRE,} {#2}, #3}
\newcommand{\yphl}[3]{ #1, {Phys. Lett.,} {#2}, #3}
\newcommand{\yptrs}[3]{ #1, {Phil. Trans. Roy. Soc.,} {#2}, #3}
\newcommand{\ymn}[3]{ #1, {MNRAS,} {#2}, #3}
\newcommand{\ynat}[3]{ #1, {Nat,} {#2}, #3}
\newcommand{\ysci}[3]{ #1, {Sci,} {#2}, #3}
\newcommand{\ypr}[3]{ #1, {Phys. Rev.} {#2}, #3}
\newcommand{\tpr}[2]{ ~#1~ {Phys. Rev. }{\bf #2} (in prep.)}
\newcommand{\spr}[2]{ ~#1~ {Phys. Rev. }{\bf #2} (submitted)}
\newcommand{\sprt}[2]{ #1, {Phys. Rep. } (submitted, #2)}
\newcommand{\ppr}[2]{ ~#1~ {Phys. Rev. }{\bf #2} (in press)}
\newcommand{\yicarus}[3]{ #1, {Icarus,} {#2}, #3}
\newcommand{\yspd}[3]{ #1, {Sov. Phys. Dokl.,} {#2}, #3}
\newcommand{\yjcp}[3]{ #1, {J. Comput. Phys.,} {#2}, #3}
\newcommand{\yjour}[4]{ #1, {#2}, {#3}, #4}
\newcommand{\yprep}[2]{ #1, {\sf #2}}
\newcommand{\ybook}[3]{ #1, {#2} (#3)}
\newcommand{\yproc}[5]{ #1, in {#3}, ed. #4 (#5), #2}
\newcommand{\pproc}[4]{ #1, in {#2}, ed. #3 (#4) (in press)}
\newcommand{\ppp}[1]{ #1, {Phys. Plasmas} (in press)}
\newcommand{\sapj}[1]{ #1, {ApJ} (submitted)}
\newcommand{\sana}[1]{ #1, {A\&A} (submitted)}
\newcommand{\san}[1]{ #1, {AN} (submitted)}
\newcommand{\tprl}[1]{ #1, {Phys. Rev. Lett.,} (in prep.)}
\newcommand{\sprl}[1]{ #1, {Phys. Rev. Lett.,} (submitted)}
\newcommand{\pprl}[1]{ #1, {Phys. Rev. Lett.,} (in press)}
\newcommand{\sjfm}[1]{ #1, {JFM} (submitted)}
\newcommand{\sgafd}[1]{ #1, {Geophys. Astrophys. Fluid Dyn.} (submitted)}
\newcommand{\pgafd}[1]{ #1, {Geophys. Astrophys. Fluid Dyn.} (in press)}
\newcommand{\tana}[1]{ #1, {A\&A} (to be submitted)}
\newcommand{\smn}[1]{ #1, {MNRAS} (submitted)}
\newcommand{\pmn}[1]{ #1, {MNRAS} (in press)}
\newcommand{\papj}[1]{ #1, {ApJ} (in press)}
\newcommand{\papjl}[1]{ #1, {ApJL} (in press)}
\newcommand{\sapjl}[1]{ #1, {ApJL} (submitted)}
\newcommand{\pana}[1]{ #1, {A\&A} (in press)}
\newcommand{\pan}[1]{ #1, {AN} (in press)}
\newcommand{\pjour}[2]{ #1, {#2} (in press)}
\newcommand{\sjour}[2]{ #1, {#2} (submitted)}
%
\newcommand\B{B}
\newcommand\E{E}
\newcommand\aaa{a}
\newcommand{\crit}{_\mathrm{c}}
\newcommand{\eff}{_\mathrm{e}}
\newcommand{\Kin}{_\mathrm{K}}
\newcommand{\M}{_\mathrm{M}}
\newcommand{\etat}{\eta_\mathrm{t}}
\newcommand\deriv[2]{\displaystyle\frac{\partial #1}{\partial #2} }
\newcommand\sfrac[2]{{\textstyle{\frac{#1}{#2}}}}
\newcommand{\Vh}{U_\mathrm{h}}
\newcommand{\Veff}{U}
\newcommand{\Beq}{\B_\mathrm{eq}}
\newcommand{\Rev}{R_\Veff}
\newcommand{\ld}{l_\mathrm{d}}
%
%
\newcommand{\cmcube}{\,{\rm cm^{-3}}}
\newcommand{\Jy}{\,{\rm Jy}}
\newcommand{\Jyb}{\,{\rm Jy/beam}}
\newcommand{\mJy}{\,{\rm mJy}}
\newcommand{\mJyb}{\,{\rm mJy/beam}}
\newcommand{\mG}{\,{\rm mG}}
\newcommand{\mkG}{\,\mu{\rm G}}
\newcommand{\MHz}{\, {\rm MHz}}
\newcommand{\Msol}{\,{\rm M_\odot}}
\newcommand{\p}{\,{\rm pc}}
\newcommand{\radm}{\,{\rm rad\,m^{-2}}}

\title{Galactic dynamo and helicity losses through fountain flow}

\author{
Anvar Shukurov\inst{1,3}
\and
Dmitry Sokoloff\,\inst{2,3}
\and
Kandaswamy Subramanian\inst{3}
\and
Axel Brandenburg\,\inst{4}
}

\offprints{anvar.shukurov@ncl.ac.uk}

\authorrunning{A.~Shukurov, D.~Sokoloff,  K.~Subramanian, \& A.~Brandenburg}
\titlerunning{Galactic dynamo and helicity losses}

\institute{
School of Mathematics and Statistics, University of Newcastle, Newcastle
upon Tyne, NE1 7RU, UK 
\and
Department of Physics, Moscow
University, 119992 Moscow, Russia 
\and
Inter-University Centre for Astronomy and
        Astrophysics,  Post Bag 4, Ganeshkhind, Pune 411 007, India 
\and
Nordita, Blegdamsvej 17, DK-2100 Copenhagen \O, Denmark 
}

\date{Received ...; accepted ...}

\abstract{}{
Nonlinear behaviour of galactic dynamos is studied, allowing for
magnetic helicity removal by the galactic fountain flow.}{
A suitable advection speed is estimated, and
a one-dimensional mean-field dynamo model with
dynamic $\alpha$-effect is explored.
}{
It is shown that the galactic fountain flow is efficient in removing
magnetic helicity from galactic discs. This alleviates the constraint on the
galactic mean-field dynamo resulting from magnetic helicity
conservation and thereby allows the mean
magnetic field to saturate at a strength
comparable to equipartition with the turbulent kinetic energy.
}{}


\maketitle
\section{Introduction}
The major controversy in
mean-field dynamo theory
is related to its nonlinear
form relevant when the
initial exponential growth of the
large-scale magnetic field saturates and the field
reaches statistical equilibrium. The core of the problem is the effect  of
the small-scale (turbulent) magnetic field on the evolution of the
large-scale (mean) magnetic field. It has been argued (Vainshtein \& Cattaneo
\cite{VC92},
Cattaneo \& Hughes \cite{CH96}) that the Lorentz force
due to the rapidly growing small-scale magnetic field
$\vec b$ can make the large-scale dynamo action inefficient
so that it produces
only a negligible mean field $\meanBB$, with $\meanBB^2\simeq
\Rm^{-1}\overline{\bb^2}$, where $\Rm\ (\gg1)$ is the magnetic Reynolds number.
Here and below, overbars denote averaged quantities.
In particular, the $\alpha$-effect (a key ingredient of
the mean-field dynamo) is catastrophically quenched well before the
large-scale magnetic field can be amplified to
the strength observed in astrophysical objects,
$\meanB\simeq b$.

The suppression of the $\alpha$-effect can be
a consequence of the conservation of magnetic helicity in a
medium of high electric conductivity
(see Brandenburg \& Subramanian \cite{BS05a} for a review).
In a closed system, magnetic helicity can only evolve on the Ohmic
time scale which is proportional to $\Rm$;
in galaxies, this time scale by far exceeds the Hubble time.
Since the large-scale magnetic field necessarily has non-zero helicity
in each hemisphere through the mutual linkage of poloidal and toroidal fields, the
dynamo also has to produce small-scale helical magnetic fields with
the opposite sign of magnetic helicity.
Unless the small-scale magnetic field can be transported out of the system,
it quenches the $\alpha$-effect together with the mean-field dynamo.

Blackman \& Field (\cite{BF00}) first suggested that the losses of the
small-scale magnetic helicity through the boundaries of the dynamo region can
be essential for mean-field dynamo action. Such a helicity flux can result
from the anisotropy of the turbulence combined with large-scale velocity shear
(Vishniac \& Cho \cite{VC01}, Subramanian \& Brandenburg \cite{SB04}) or the
non-uniformity of the $\alpha$-effect (Kleeorin et al.\ \cite{KMRS00}). The
effect of the Vishniac-Cho flux has already been confirmed in simulations,
allowing the production of significant fields, $\meanB\simeq b$
(Brandenburg \cite{B05}).

Here we suggest another simple mechanism where the advection of small-scale
magnetic fields (together with the associated magnetic helicity) away from the
dynamo region allows healthy mean-field dynamo action. This effect naturally
arises in spiral galaxies where magnetic field is generated in the multi-phase
interstellar medium. The mean magnetic field is apparently produced by the
motions of the warm gas (\S4.3 in Beck et al.\ \cite{BBMSS96}), which is in a
state of (statistical) hydrostatic equilibrium with a scale height of
$h\simeq0.5\kpc$ (e.g.\ Korpi et al.\ \cite{K99}).  However, some of the gas
is heated by supernova explosions producing a hot phase whose isothermal scale
height is 3\,kpc.
The hot gas leaves the galactic disc, dragging along the small-scale part of
the interstellar magnetic field. Thus, the disc-halo connection in spiral
galaxies represents a mechanism of transport of small-scale magnetic fields
and small-scale magnetic helicity from the dynamo active disc to the galactic
halo. As we show here, this helps to alleviate the catastrophic
$\alpha$-quenching under realistic parameters of the interstellar medium.

\section{The disc-halo connection in spiral galaxies}	\label{DHI}
The hot gas produced by supernovae cannot be confined to the galactic disc and
therefore it is involved in systematic vertical motions. The gas flows to the
halo where it cools and contracts at a height of several kiloparsecs and then
falls down to the disc in the form of cold, relatively small clouds, forming
what is known as the galactic fountain (Shapiro \& Field \cite{SF76}, Bregman
\cite{Br}). The initial
vertical velocity of the hot gas is $u_z=100$--$200\kms$ (Norman \&
Ikeuchi \cite{NI89}), Kahn \& Brett \cite{KB93}, Avillez \& Berry \cite{AB01}).

The fountain flow drives gas mass flux from the disc, $\dot M=2\pi
R^2\rho_\mathrm{h} f u_z$ through both surfaces, where $f$ is the area
covering factor of the hot gas, $R\simeq15\kpc$ is the radius of the galactic
disc with vigorous supernova activity, and
$\rho_\mathrm{h}\simeq1.7\times10^{-27}\,\mbox{g}\cm^{-3}$ is the hot gas
density. A lower estimate of the area covering factor is given by the volume
filling factor of the hot gas at the disc midplane, $f=0.2$--0.3 (e.g.\ Korpi
et al.\ \cite{K99}), because the scale height of the galactic disc is
comparable to the size of the hot cavities. This yields $\dot M\simeq
1.5\,M_\odot\yr^{-1}$ (cf.\ Norman \& Ikeuchi \cite{NI89} who obtain $\dot
M=0.3$--$3\,M_\odot\yr^{-1}$).

The dynamo model discussed below refers to quantities averaged over scales
exceeding the size of the hot cavities, and so it treats the multi-phase
interstellar medium in an averaged manner. Then it is appropriate to introduce
an effective fountain speed as the one that drives the same mass flux as above
by advecting the diffuse interstellar gas at its mean density,
$\rho=1.7\times10^{-25}\,\mbox{g}\cm^{-3}$
(i.e.\ number density of $0.1\cm^{-3}$)
\begin{equation}                \label{Veff}
\Veff_0=f u_z\,\frac{\rho_\mathrm{h}}{\rho}\simeq1\mbox{--}2\kms\;,
\end{equation}
and less for larger $\rho$ (say, 0.3--$0.7\kms$ for the gas number density of
$0.3\cm^{-3}$). Given the uncertainty of the magnitude of $f$ (e.g.,
$f\simeq0.1$ in the model of  Norman \& Ikeuchi \cite{NI89}), the range
$\Veff_0=0.2$--$2\kms$ seems to be plausible.

The hot gas that leaves the galactic disc carries magnetic fields of
scales smaller than the size of the hot cavities
(0.1--1\,kpc); these are mostly turbulent magnetic fields
(of scales $\la100\p$). The time
scale of the removal of the small-scale magnetic fields is of order
$h/\Veff_0\simeq 5\times10^8\yr$ (with $h\simeq0.5\kpc$ the scale height of
the warm gas layer which hosts the mean-field dynamo), which is comparable to
the turbulent diffusion time of the mean field, $h^2/\eta_{\rm t}
\simeq8\times10^8\yr$, with $\etat=10^{26}\cm^2\s^{-1}$ the turbulent
magnetic diffusivity. Hence,  the leakage of the small-scale magnetic helicity
produced by the galactic fountain can significantly affect the mean-field
dynamo.

\section{Magnetic helicity balance with fountain flow} \label{MFD}
In the following we use,
wherever appropriate,
lower case characters to denote random, small scale quantities
defined as the departure from the corresponding mean, e.g.\
$\aaaa=\AAA-\meanAA$, $\jj=\JJ-\meanJJ$ and $\uu=\UU-\meanUU$  for the
vector potential, current density and velocity, respectively.
A gauge-invariant magnetic helicity {\em density\/} $\chi$ can be defined for
$\bb$ as the density of correlated links of $\bb$ using Gauss's linkage
formula (Subramanian \& Brandenburg \cite{SB05}). Such a definition can be
introduced for random magnetic fields if their correlation length is finite
and much smaller than the system size. The helicity density
$\overline{\aaaa\cdot\bb}$ defined with the Coulomb gauge, $\nab\cdot\aaaa=0$,
is only
approximately the same as $\chi$. In the presence of a helicity flux $\FF$,
the magnetic helicity density for the small-scale field evolves as
(Subramanian \& Brandenburg \cite{SB05})
\EQ
\frac{\partial \chi}{\partial t} + \nab\cdot\FF
= -2\meanEMF\cdot\meanBB-2\eta\overline{\jj\cdot\bb},
\label{finhel}
\EN
where $\meanEMF = \overline{\uu \times \bb}$ is the turbulent emf and $\eta$
is the Ohmic magnetic diffusivity.
(We use units where $\JJ=\nabla\times\BB$.) In
the steady state, Eq.~(\ref{finhel}) yields $\meanEMF\cdot\meanBB =
-\half\nab\cdot\FF -\eta\overline{\jj\cdot\bb}$. Without a helicity flux,
$\meanEMF\cdot\meanBB = -\eta\overline{\jj\cdot\bb}$ follows, which vanishes
as $\eta \to 0$ for any reasonable spectrum of current helicity
$\overline{\jj\cdot\bb}$ so that $\overline{\jj\cdot\bb}\propto
\eta^{-\kappa}$ with $\kappa<1$. In the presence of helicity fluxes, however,
$\meanEMF\cdot\meanBB$ needs not be catastrophically quenched as $\eta\to0$.

We study the effects of the simplest contribution to the flux,
\[
\FF = \chi\meanUU\,,
\]
which arises from the net effect of advection by upward and downward flows
(cf.\ Subramanian \& Brandenburg \cite{SB05}) dominated by the upward
flow in the case of galactic fountains.
The overall effect of the resulting helicity advection from the
disc of the galaxy is not immediately obvious since the flow can
also remove the mean magnetic field from the dynamo-active region.
To reveal the net effect of the fountain flow on the dynamo we solve
the helicity equation simultaneously with the mean-field dynamo equation,
\EQ
{\partial\meanBB\over\partial t}=
\nab\times(\meanUU\times\meanBB+\meanEMF-\eta\meanJJ).
\label{fullset1flux}
\EN
We adopt
$\meanEMF=\alpha\meanBB -\eta_{\rm t}\meanJJ$
(assuming isotropic turbulence) and take
$\alpha = \alpha_\mathrm{K}+\alpha_{\rm m}$
(following Pouquet et al.\ \cite{P76}), where
$\alpha_{\rm K}$
represents the kinetic $\alpha$-effect,
$\alpha_{\rm m} = \sfrac13\rho^{-1}\overline{\tau\jj\cdot\bb}$ is the magnetic
contribution to the $\alpha$-effect,
and $\eta_{\rm t} =\sfrac13 \overline{\tau \uu^2}$.

The dynamics of $\alpha_{\rm m}$ is controlled by
Eq.~(\ref{finhel}).
We argue that the main contribution to
$\alpha_{\rm m}$ comes from the integral scale of the turbulence $l_0=2\pi/\ko$.
For Kolmogorov turbulence, we have the following spectral scalings:
$\jj_k\cdot\bb_k\propto k b_k^2\propto
k^{1/3}$ and $\tau_k\propto k^{-2/3}$, so that $\alpha_\mathrm{m}\propto k^{-1/3}$.
Moreover, numerical results of Brandenburg \& Subramanian (\cite{BS05b}) indicate that
even $\overline{\jj\cdot\bb}$ is dominated by the larger scales.
This justifies the estimate
\[
\alpha_{\rm m} \simeq\sfrac13\tau \ko^2 \,\frac{\chi}{\rho}\,.
\]
With
$B_{\rm eq}^2\equiv\rho \mean{u^2}$ and $\Rm=\eta_{\rm t}/\eta$,
we
then
rewrite Eq.~(\ref{finhel}) as
(cf.\ Blackman \& Brandenburg \cite{BB02}, Subramanian \cite{S02}),
\begin{equation}
{\partial\alpha_{\rm m}\over\partial t}=-2\eta_{\rm t} \ko^2\left(
{\meanEMF\cdot\meanBB\over B_{\rm eq}^2}
+{\alpha_{\rm m}\over \Rm}\right)
-\nab\cdot\left(\alpha_{\rm m}\meanUU\right).
\label{fullset2flux}
\end{equation}

Since galactic discs are thin, it suffices to consider a
one-dimensional model, retaining only the $z$-derivatives of the variables
(Ruzmaikin et al.\ \cite{RSS88}).
In terms of cylindrical coordinates $(r,\phi,z)$, we adopt
a mean flow consisting of rotation (shear) and vertical advection,
with $\meanUU = (0, \meanU_\phi,\meanU_z)$, where
$\meanU_z=\Veff_0 z/h$ within the disc.
We use $\alpha_{\rm K}=\alpha_0\,z/h$.
The mean-field dynamo equation, in cylindrical coordinates, becomes
\EQ \label{mfBr}
{\partial\meanB_r\over\partial t}= -{\partial\over\partial
z}\left(\meanU_z\meanB_r+\meanemf_\phi\right)
+\eta{\partial^2\meanB_r\over\partial z^2},
\EN
\EQ \label{mfBphi}
{\partial\meanB_\phi\over\partial t}=
-{\partial\over\partial z}\left(\meanU_z\meanB_\phi-\meanemf_{r}\right)
+\eta{\partial^2\meanB_\phi\over\partial z^2}
+q\Omega_0\meanB_r,
\EN
where $q\approx-1$ for a flat rotation curve, $\Omega\propto r^{-1}$,
and $\Omega_0$ is the local rotation rate.
We solve Eqs~(\ref{fullset2flux})--(\ref{mfBphi}) with the boundary conditions
\EQ \label{bconds}
\meanB_r=\meanB_\phi=0
\quad\mbox{
at $z=\pm h$}.
\EN
We adopt $\alpha_\mathrm{m}=0$ at $t=0$, and a symmetry condition
$\alpha_\mathrm{m}=0$ at $z=0$ follows automatically from the symmetry of $\meanBB$; the two
conditions result in a unique solution of Eq.~(\ref{fullset2flux}).
(There is no need to be specify separately $\alpha_\mathrm{m}$ at $z=\pm h$.)
We introduce dimensionless numbers
\EQ
C_U={\Veff_0 \over\eta_{\rm t}k_1},\quad
C_\Omega={\Omega_0\over\eta_{\rm t}k_1^2},\quad
C_\alpha={\alpha_0 \over\eta_{\rm t}k_1},
\EN
where $k_1=\pi/h$. For the fiducial values $h=500\p$, $\Veff_{0}=0.2$--$2\kms$,
$\Omega_0=25\km\s^{-1}\kpc^{-1}$, $\alpha_0=0.5\kms$ and
$\etat=10^{26}\cm^2\s^{-1}$, we have
$C_U=0.1\mbox{--}1,\ C_\Omega\approx-2$ and $C_\alpha\approx0.8$.
We use $k_0/k_1=h/l_0=5$ as appropriate for the galactic disc,
and present our results in the units of
\[
B_\mathrm{eq}\approx5\mkG\quad
\mbox{and}\quad
(\etat k_1^2)^{-1}\approx8\times10^7\yr
\]
for magnetic field and time, respectively (where gas number density of
$1\cm^{-3}$ and the turbulent velocity of $10\kms$
has been used in $B_\mathrm{eq}$).

\section{Results}
\begin{figure}[t!]\begin{center}
\includegraphics[width=.98\columnwidth,clip]{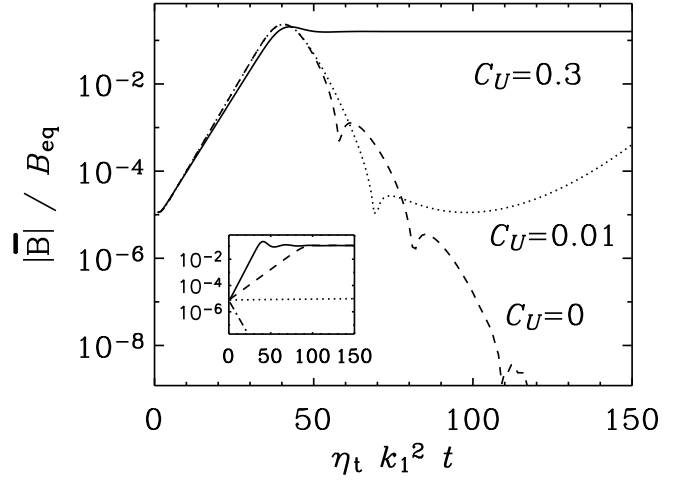}
\end{center}\caption[]{
Evolution of the field strength at $z=0$ obtained by solving
Eqs~(\ref{fullset2flux})--(\ref{bconds})
with vertical advection (solid line, $C_U=0.3$)
and without it (dashed line, $C_U=0$), for
$C_\Omega=-2$, $C_\alpha=1$ and $\Rm=10^5$. The dynamo
is neutrally stable at
$C_\alpha=0.26$ for $\C_U=0.3$ and $C_\Omega=2$.
The dotted curve, obtained for $C_U\ll1$, shows that even weak advection can
affect the long-term evolution of magnetic field.
For $\C_U=0$, nonlinear effects make the $\alpha$ profile flatter
at small $|z|$; this causes an oscillatory decay of the field.
The inset shows similar results for $\C_U=0.1$ (solid),
$1.5$ (dashed), $2$ (dotted) and $3$ (dash-dotted).
}\label{axsinz}
\end{figure}

We compare in \Fig{axsinz} the evolution, with and without the advective flux,
of $\bra{\meanBB^2}/B_{\rm eq}^2$,
where $\bra{\meanBB^2}$ is the mean square of the large-scale magnetic field over
the range $|z|<h$.
The solutions are non-oscillatory and have quadrupolar parity;
they grow for $C_\alpha>0.2$ if $C_\Omega=-2$ and $C_U=0$.
For the fiducial parameters, the $e$-folding time of the dynamo
is about $3\times10^8\yr$ for $C_U\la0.3$.
In the absence of  an advective flux, the
initial growth of magnetic field is
catastrophically quenched and the large-scale magnetic field decreases at about
the same rate as it grew.
The initial growth occurs while the current helicity builds up to
cancel the kinetic $\alpha$-effect.
However, even a modest advective flux ($C_U=0.1$) compensates the
catastrophic quenching of the dynamo and the mean field energy density
stays steady at about $10^{-2} B_{\rm eq}^2$,
which corresponds to the field strength of about 10\%\
of the equipartition value.

We illustrate
in the inset of \Fig{axsinz} the effect of varying the strength of
the advection velocity on the dynamo.
The steady-state strength of the mean field grows with
$\Veff_0$ as the advection strength increases to about $C_U=0.3$.
For stronger advection, $C_U\simeq0.5$, the mean field
initially grows slower but still attains a steady state
strength slightly exceeding $0.1B_\mathrm{eq}$. Stronger advection,
$C_U > 1$, affects the dynamo adversely since the mean field is removed too
rapidly from the dynamo active region.
A good compromise between rapid growth and large saturation field strength
is reached for $C_U\approx0.1$,
which is close to the values expected for spiral galaxies.

\begin{figure}[t!]\begin{center}
\centerline{\includegraphics[width=.95\columnwidth,clip]{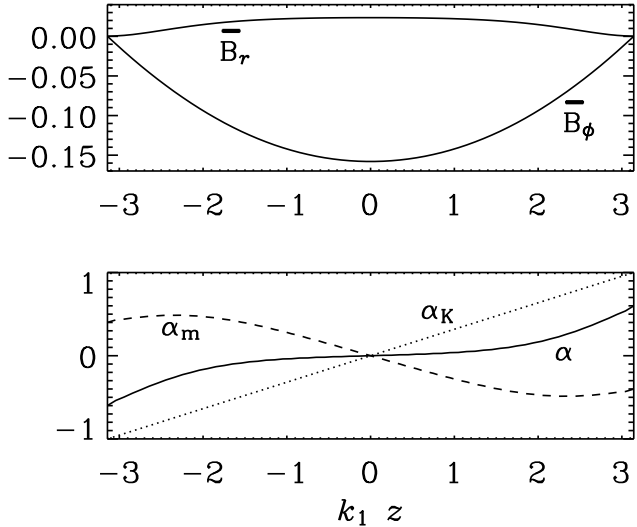}}
\end{center}\caption[]{
Plots of $\meanB_\phi$ and $\meanB_r$ (upper panel)
and $\alpha$ (solid), $\alpha_{\rm m}$ (dashed) and $\alpha_\mathrm{K}$
(dotted) (lower panel)
versus $z$ in the steady state
for
$C_U=0.3$, $C_\Omega=-2$, $C_\alpha=1$, $R_{\rm m}=10^5$, and
$k_1=\pi/h$.
}\label{pcomp_adflux34cross}
\end{figure}

Modest advection does not noticeably affect the spatial distribution of magnetic field.
The profiles of $\meanB_\phi$ and $\meanB_r$ shown in
\Fig{pcomp_adflux34cross} for the steady state do not differ much
from
solutions
of the kinematic dynamo equations (cf.\ Ruzmaikin et
al.\ \cite{RSS88}).
The corresponding profiles of $\alpha$ and $\alpha_{\rm m}$ shown
in the lower panel, indicate that the suppression of $\alpha$ in stronger
near the disc midplane (at small $z$), where magnetic field is stronger.

The steady-state strength of $\meanBB$, can be estimated
from magnetic helicity conservation.
Averaging \Eq{fullset2flux}, with $\partial\alpha_\mathrm{m}/\partial t=0$, over
$z$ on $0<z<h$ (with the mean denoted with angular brackets) yields
\begin{equation}\label{balance}
\frac{\bra{\meanEMF\cdot\meanBB}}{B_{\rm eq}^2} +
\frac{\bra{\alpha_{\rm m}}}{R_{\rm m}}
+\frac{\alpha_{\rm m}(h) \meanU_z(h)}{2\eta_{\rm t} k_0^2 h}=0,
\end{equation}
since $\Veff_z(0) = 0$.
For the adopted boundary conditions
(\ref{bconds}) and $|\meanB_z|\ll|\meanB_{r}|,|\meanB_\phi|$ within the disc,
it is meaningful to calculate magnetic helicity within the disc only.
Advection
does not affect helicity conservation, which
then implies
$\bra{\meanJJ\cdot\meanBB}
\simeq k_1^2 \bra{\meanAA\cdot\meanBB}
= -k_1^2 \chi$.
Given that
$\meanEMF\cdot\meanBB/B_\mathrm{eq}^2\simeq(\alpha_\mathrm{K}
+\alpha_\mathrm{m})(\meanB/B_\mathrm{eq})\mbox{}^2
+\alpha_\mathrm{m}(k_1/k_0)^2$, and assuming that
$\alpha_\mathrm{K}+\alpha_\mathrm{m}=\alpha_\mathrm{c}$, where
$\alpha_\mathrm{c}$ corresponds to a marginally stable dynamo,
Eq.~(\ref{balance}) yields the following estimate of magnetic field strength
in the steady state:
\begin{equation}\label{steady}
\meanB\simeq B_\mathrm{eq}
\,\frac{k_1}{k_0}\left(\frac{D}{D_\mathrm{c}}-1\right)^{1/2}\,C^{1/2},
\end{equation}
where $D=C_\alpha C_\Omega$ is the dynamo number and $D_\mathrm{c}$ is its
marginal value, and $C=1+C_U/(2\pi)+k_0^2/(\Rm k_1^2)=O(1).$
With $k_1/k_0\simeq5$ and $D\simeq D_\mathrm{c}$,
this implies
$\meanB\simeq 0.2 B_\mathrm{eq}$ in rough agreement with the numerical solution.
We note that Eq.~(\ref{steady}) does not
apply for $|D|\gg|D_\mathrm{c}|$ because our arguments are only valid for
weakly nonlinear regimes.

\section{Discussion}
The vertical advection of magnetic helicity by galactic fountain flow
resolves straightforwardly the
controversy of nonlinear mean-field galactic dynamos.
For $C_U=0$, the mean magnetic field does initially reach a level consistent
with Eq.~(\ref{steady}), but then rapidly decays to negligible values
(see Brandenburg \& Subramanian \cite{BS05c}).
The essential role of the advection is to provide the
system with an opportunity to reach and then maintain
the steady state (\ref{steady}) within the galactic lifetime.

These conclusions follow from Fig.~\ref{axsinz}, where
moderate advection drastically changes the mean field levels achievable at
$t\la10^{10}\yr$ and prevents catastrophic quenching of the dynamo.
Excessive advection, however, hinders the dynamo as it
removes the mean field from the dynamo active region.

The steady-state strength of the mean magnetic field obtained in our model is
of order $(0.1$--$0.2)\,B_\mathrm{eq}\simeq0.5$--$1\mkG$, which is a factor of
several weaker than what is observed; we made no attempt here to refine the model.
What is important, the mechanism suggested here resolves the problem of
catastrophic quenching of the dynamo.

The applicability of the vacuum boundary conditions (\ref{bconds}) to a system
with gas outflow from the disc can be questionable. Analysis of dynamo models
with outflow (Bardou et al.\ \cite{BR01}) suggests that reasonable changes to
the boundary conditions do not affect the dynamo too strongly, but this aspect
of our model should be further explored.

We have neglected the intrinsic difference of the
behaviours of the mean and turbulent magnetic fields near the disc surface.
Since the horizontal size of the hot cavities is larger than the scale of the
turbulent magnetic field but smaller than the scale of the mean magnetic field,
the Lorentz force can resist the advection of the mean field more efficiently
than that of the turbulent field.
Furthermore, large-scale magnetic field loops drawn out by the fountain flow
can be detached from the parent magnetic lines by reconnection, so that the
flow will carry mostly small-scale fields. Therefore, a more plausible (albeit
less conservative) model would include advection of the small-scale (but not
the large-scale) magnetic field. In such a model, the effect discussed here
can be even better pronounced.

Brandenburg et al.\ (\cite{BMS95}) argue that the fountain flow can
transport the large-scale magnetic field into the halo by topological
pumping if the hot gas forms a percolating cluster in the disc and if
the turbulent magnetic Reynolds number $C_U$ in the fountain flow
exceeds $\simeq20$; our estimate given below is
$C_U\la 1$, and we
expect that the small-scale magnetic fields are removed from the disc
more efficiently than the mean field.

The idea that advection of
small-scale magnetic fields can help the galactic dynamo may be
more robust than our particular model of dynamo quenching
that involves the magnetic $\alpha$-effect.
For example, if the dynamo coefficients are quenched due to
the
suppression of Lagrangian chaos by the small-scale magnetic fields (Kim \cite{Kim99}),
their advection out of the galaxy
will still allow the dynamo to operate efficiently.


\begin{acknowledgements}
We are grateful to David Moss for helpful comments.
KS acknowledges the hospitality of NORDITA.
This work was supported by the Royal Society, RFBR grant 04-02-16094 and INTAS
grant 2021.
\end{acknowledgements}


\end{document}